\documentclass[12pt]{article}
\usepackage{amsmath}
\input{epsf} 
\newcommand\as{\alpha_{\mathrm{S}}} 
\newcommand\f[2]{\frac{#1}{#2}} 
\def\nn{\nonumber}
\def\ltap{\raisebox{-.4ex}{\rlap{$\,\sim\,$}} \raisebox{.4ex}{$\,<\,$}}

\begin{document}
\begin{titlepage}
\renewcommand{\thefootnote}{\fnsymbol{footnote}}
\begin{flushright}
hep-ph/0209302
 \end{flushright}
\par \vspace{10mm}

\begin{center}
{\Large \bf
Direct Higgs production at hadron colliders
\footnote{Talk given at the 31$^{\rm st}$ International Conference on High Energy Physics, ICHEP, Amsterdam, July 2002}
\footnote{Work done in collaboration with S.~Catani, D.~de~Florian and P.~Nason}
\\[1.ex]
}

\end{center}
\par \vspace{2mm}
\begin{center}
{\bf Massimiliano Grazzini${}^{(a,b)}$}\\

\vspace{5mm}

${}^{(a)}$Dipartimento di Fisica, Universit\`a di Firenze, 
I-50019 Sesto Fiorentino, Florence, Italy\\

${}^{(b)}$INFN, Sezione di Firenze, I-50019 Sesto Fiorentino, Florence, Italy

\end{center}

\par \vspace{2mm}
\begin{center} {\large \bf Abstract} \end{center}
\begin{quote}
\pretolerance 10000

We consider QCD corrections to Higgs boson production through gluon-gluon fusion in hadron collisions. 
We compute the effects of soft-gluon emission to all orders. We present numerical results at the Tevatron and the LHC. 

\end{quote}

\vspace*{\fill}
\begin{flushleft}
September 2002 

\end{flushleft}
\end{titlepage}

\setcounter{footnote}{1}
\renewcommand{\thefootnote}{\fnsymbol{footnote}}

The most important mechanism for SM Higgs boson production at hadron colliders
is gluon--gluon fusion through a heavy (top) quark loop.
NLO QCD corrections to this process are known to be large
\cite{Dawson:1991zj,Djouadi:1991tk,Spira:1995rr}: their effect increases
the LO inclusive cross section by about 80-100$\%$. 
Recently, the calculation of the NNLO corrections has been completed in the large-$M_t$ approximation
\cite{Catani:2001ic,Harlander:2001is,Catani:2001cr,Harlander:2002wh,Anastasiou:2002yz}.
This approximation has been shown to work well at NLO provided the exact dependence on the mass $M_t$ of the top quark is retained in the LO result.

\begin{figure}[htb]
\begin{center}
\begin{tabular}{c}
\epsfxsize=12truecm
\hskip -0.5cm\epsffile{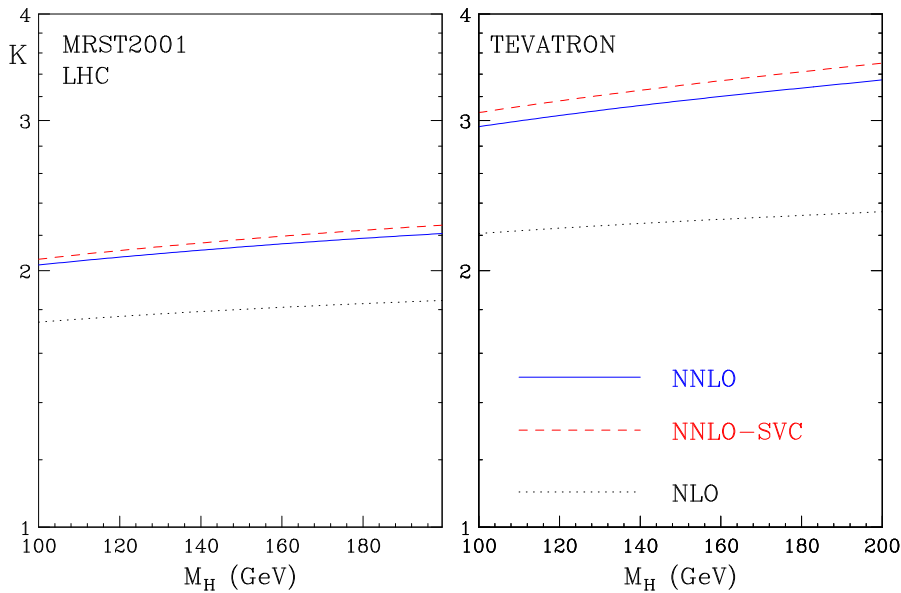}
\end{tabular}
\end{center}
\caption{\label{fig:nnlo}{\em K factors at the Tevatron and the LHC}.}
\end{figure}

In Fig.~\ref{fig:nnlo} we show the NNLO effect with respect to NLO
at the Tevatron and the LHC, as a function of
the mass $M_H$ of the SM Higgs boson.
We use the MRST2001 set \cite{mrst2001},
which includes approximated NNLO parton distributions,
with $\as$ consistently evaluated at one-loop, two-loop,
three-loop order for the LO, NLO, NNLO results, respectively.
The factorization and renormalization scales $\mu_F$ and $\mu_R$ are fixed to $M_H$.
The solid line is the full NNLO result \cite{Anastasiou:2002yz}
while the dashed line is the result including only soft and virtual plus leading collinear contributions (SVC) \cite{Catani:2001ic}.
We see that the NNLO effect is moderate 
at the LHC: in the case of a light Higgs, the $K$-factor is about $2.1$--$2.2$,
corresponding to an increase of about 20$\%$ with respect to NLO.
The NNLO effect is more sizeable
at the Tevatron where $K\sim 3$, the increase being
of about $40 \%$ with respect to NLO.
Fig.~\ref{fig:nnlo} shows that
the SVC approximation
works remarkably well, the differences
with respect to the full result being only
about $2\%$ at the LHC and $4\%$ at the Tevatron.
Thus the bulk of the NNLO contributions
is due to soft and collinear radiation
\cite{Catani:2001ic,Harlander:2001is,Catani:2001cr},
which factorizes from the heavy-quark loop,
whereas the hard radiation \cite{Harlander:2002wh} gives only a very small 
effect.
The dominance of soft and collinear radiation
has two important consequences.
First, it can be considered as a justification of the use
of the large $M_t$-approximation at NNLO.
Second, it suggests that
multiple soft-gluon emission beyond NNLO can give a relevant effect.
Here we discuss the effects of resummation of soft (Sudakov) emission
to all orders \cite{inprep}.

The cross section ${\hat \sigma}_{gg}$ for the partonic subprocess $gg\to H+X$ 
at the centre--of--mass energy ${\hat s}=M_H^2/z$ can be written as
\begin{equation}
\label{spart}
{\hat \sigma}_{gg}({\hat s},M_H^2) = \sigma_0 \;z \;G_{gg}(z) \;,
\end{equation}
where $M_H$ is the Higgs mass, $\sigma_0$ is the Born-level cross section and
$G_{gg}$ is the perturbatively computable coefficient function.
Soft-gluon resummation
is performed \cite{Catani:1996yz} in Mellin (or $N$-moment) space 
($N$ is the moment variable conjugate to $z$).
The {\em all-order} resummation formula for the
coefficient function ${G}_{gg}$ is \cite{Catani:2001ic,Kramer:1996iq}: 
\begin{equation}
\label{resformula} 
{G}_{gg, \, N}^{{\rm (res)}} =\as^2(\mu_R^2)\,
C_{gg}(\as(\mu^2_R),M_H^2/\mu^2_R;M_H^2/\mu_F^2)\;
\Delta_{N}^{H}(\as(\mu^2_R),M_H^2/\mu^2_R;M_H^2/\mu_F^2)\;.
\end{equation}
The function $C_{gg}(\as)$ contains all the terms that are 
constant in the large-$N$ limit,
produced by hard virtual contributions and non-logarithmic soft corrections. 
It can be computed as a power series expansion in $\as$ as 
\begin{equation}
\label{coef}
C_{gg}(\as(\mu^2_R),M_H^2/\mu^2_R;M_H^2/\mu_F^2) =
1 + \sum_{n=1}^{+\infty} 
\left( \frac{\as(\mu^2_R)}{\pi} \right)^n 
C_{gg}^{(n)}(M_H^2/\mu^2_R;M_H^2/\mu_F^2) \;,
\end{equation}
where the perturbative coefficients $C_{gg}^{(n)}$ are closely related
to those of the $\delta(1-z)$ contribution to $G_{gg}(z)$. 
The radiative factor $\Delta_{N}^{H}$ embodies the large logarithmic terms
due to soft-gluon radiation.
To implement resummation, the radiative factor is expanded to a given 
logarithmic accuracy as
\begin{align} 
\label{deltannll} 
\Delta_N^{H} = 
\exp &\Big\{ \f{}{} \ln N \; g^{(1)}(\lambda) + 
g^{(2)}\!\left(\lambda,\f{M_H^2}{\mu^2_R};\f{M_H^2}{\mu_F^2}\right)\nn\\
&\; + \as(\mu^2_R)
\, g^{(3)}\!\left(\lambda,\f{M_H^2}{\mu^2_R};\f{M_H^2}{\mu_F^2}\right)  
+ {\cal O}(\as^2(\as \ln N)^k) \Big\} , 
\end{align} 
such that the functions $g^{(1)}, g^{(2)}$ and  $g^{(3)}$ respectively
collect the leading logarithmic (LL), next-to-leading logarithmic (NLL)
and next-to-next-to-leading logarithmic (NNLL) terms with respect to the
expansion parameter $\lambda=\as(\mu^2_R) \ln N$.

NLL resummation \cite{Catani:2001ic} is controlled by three perturbative 
coefficients, $A_g^{(1)}, A_g^{(2)}$ and $C^{(1)}_{gg}$.
The coefficients $A_g^{(1)}$ and $A_g^{(2)}$, which
appear in the functions $g^{(1)}$ and $g^{(2)}$, are well known
\cite{Catani:1996yz}. The coefficient $C^{(1)}_{gg}$ in Eq.~(\ref{coef}) is
extracted from the NLO result.

At NNLL accuracy three new coefficients are needed \cite{Catani:2001ic}: 
the coefficient $C_{gg}^{(2)}$ in Eq.~(\ref{coef}) and two coefficients, 
$D^{(2)}$ and $A_g^{(3)}$, which appear in the NNLL function $g^{(3)}$.
The functional form of $g^{(3)}$ was computed
in Ref.~\cite{vogtresum}.
The coefficients $D^{(2)}$ and $C_{gg}^{(2)}$ are obtained \cite{Catani:2001ic}
from the NNLO result. The coefficient $A_g^{(3)}$ is not yet fully known:
we use its exact $N_f^2$-dependence \cite{Bennett:1997ch} and the approximate
numerical estimate of Ref.~\cite{vanNeerven:2000wp}.

Finally, the dominant collinear logarithmic terms can be accounted for
by modifying the coefficient $C_{gg}^{(1)}$ in the resummation formula as
\cite{Catani:2001ic}
\begin{equation}
\label{colterm}
C_{gg}^{(1)} \rightarrow C_{gg}^{(1)} + 2 A_g^{(1)} \; \f{\ln N}{N} \;.
\end{equation}

In the following we present a preliminary study of
the resummation effect at the Tevatron and the LHC.
The hadron-level cross section is obtained by convoluting the partonic cross
section in Eq.~(\ref{spart}) with the parton distributions of the colliding 
hadrons. 
As in Fig.~\ref{fig:nnlo} we use the MRST2001 set.
The resummed calculations are always matched
to the corresponding fixed-order results, i.e. LL is matched to LO,
NLL to NLO and NNLL to NNLO.
We find that the effect of the inclusion of the collinear term in Eq.~(\ref{colterm}) is very small,
whereas the effect of the coefficient $A^{(3)}$ is completely negligible.

\begin{figure}[htb]
\begin{center}
\begin{tabular}{c}
\epsfxsize=12truecm
\hskip -0.5cm\epsffile{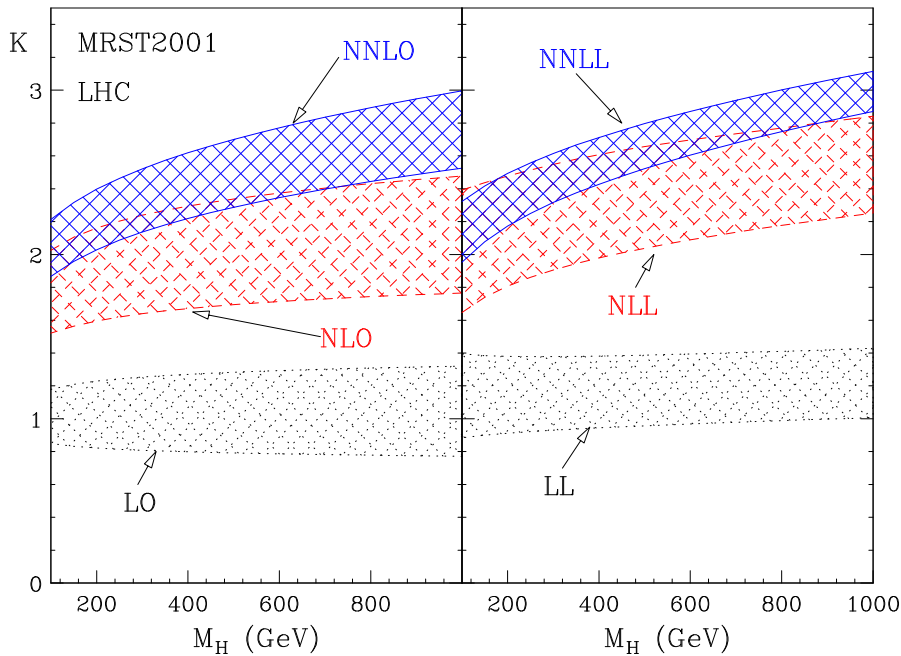}
\end{tabular}
\end{center}
\caption{\label{fig:lhc}{\em Resummed K-factors at the LHC}.}
\end{figure}

\begin{figure}[htb]
\begin{center}
\begin{tabular}{c}
\epsfxsize=12truecm
\hskip -0.5cm\epsffile{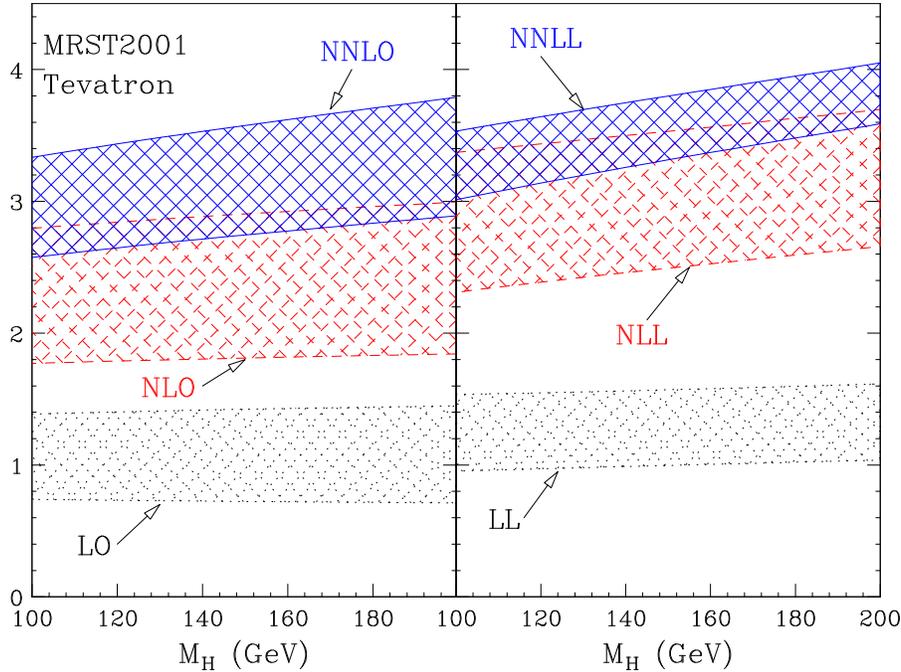}
\end{tabular}
\end{center}
\caption{\label{fig:tev}{\em Resummed K-factors at the Tevatron Run II}.}
\end{figure}

In Fig.~\ref{fig:lhc} we present our results at the LHC, by plotting
the $K$-factor, defined as the hadronic cross section as a function of $\mu_F$ and $\mu_R$, normalized to the LO result at $\mu_F=\mu_R=M_H$.
In the left side of the figure the LO, NLO and NNLO bands are shown, defined varying
$\mu_F=\mu_R$ between $0.5 M_H$ and $2M_H$.
In the right side of the figure the corresponding resummed results are plotted,
the bands being now obtained setting $\mu_F=M_H$ and letting $\mu_R$ to range between $0.5M_H$ and $2M_H$.
In both cases we have defined the bands in such a way to maximize them but avoiding completely independent scale variations such as $\mu_R=0.5 M_H$ and $\mu_F=2M_H$, by which the ratio $\mu_F/\mu_R$ would be $4$.
Different definitions of the uncertainty bands are of course possible.
We see that soft-gluon resummation gives a moderate effect, the NNLL effect being about $5-6\%$ with respect to NNLO for $M_H\ltap 200$ GeV. 
In Fig.~\ref{fig:tev} we report the analogous results at the Tevatron Run II.
The bands are defined as in Fig.~\ref{fig:lhc}.
Here the resummation effects are larger: going from NLO to NLL accuracy,
the cross section increases by 25-30$\%$.
NNLL  resummation increases the NNLO cross section by $\sim 12$-$15\%$
when $M_H$ varies in the range 100-200~GeV.
These results are not unexpected \cite{Catani:2001cr}, since at the Tevatron 
the Higgs boson is produced  
closer to threshold and the effect of multiple soft-gluon emission is more 
important. 

From these results
we conclude that the theoretical predictions for Higgs boson production at hadron colliders are under control. A more detailed discussion of the present theoretical uncertainty will be given elsewhere \cite{inprep}.


\begin{thebibliography}{90}
\bibitem{Dawson:1991zj}
S.~Dawson,
Nucl.\ Phys.\ B {\bf 359} (1991) 283.

\bibitem{Djouadi:1991tk}
A.~Djouadi, M.~Spira and P.~M.~Zerwas,
Phys.\ Lett.\ {\bf B264} (1991) 440.

\bibitem{Spira:1995rr}
M.~Spira, A.~Djouadi, D.~Graudenz and P.~M.~Zerwas,
Nucl.\ Phys.\ B {\bf 453} (1995) 17.


\bibitem{Catani:2001ic}
S.~Catani, D.~de Florian and M.~Grazzini,
JHEP {\bf 0105} (2001) 025.

\bibitem{Harlander:2001is}
R.~V.~Harlander and W.~B.~Kilgore,
Phys.\ Rev.\ D {\bf 64} (2001) 013015.

\bibitem{Catani:2001cr}
S.~Catani, D.~de Florian and M.~Grazzini,
JHEP {\bf 0201} (2002) 015.

\bibitem{Harlander:2002wh}
R.~V.~Harlander and W.~B.~Kilgore,
Phys.\ Rev.\ Lett.\  {\bf 88} (2002) 201801.

\bibitem{Anastasiou:2002yz}
C.~Anastasiou and K.~Melnikov,
hep-ph/0207004.

\bibitem{mrst2001}
A.~D.~Martin, R.~G.~Roberts, W.~J.~Stirling and R.~S.~Thorne,
Eur.\ Phys.\ J.\ C {\bf 23} (2002) 73,
Phys.\ Lett.\ B {\bf 531} (2002) 216.

\bibitem{inprep}
S.~Catani, D.~de~Florian, M.~Grazzini and P.~Nason,
in hep-ph/0204316
and preprint in preparation.

\bibitem{Catani:1996yz}
S.~Catani, M.~L.~Mangano, P.~Nason and L.~Trentadue,
Nucl.\ Phys.\ B {\bf 478} (1996) 273, and references therein.

\bibitem{Kramer:1996iq}
M.~Kramer, E.~Laenen and M.~Spira,
Nucl.\ Phys.\ B {\bf 511} (1998) 523.

\bibitem{vogtresum}
A.~Vogt,
Phys.\ Lett.\ B {\bf 497} (2001) 228.

\bibitem{Bennett:1997ch}
J.~F.~Bennett and J.~A.~Gracey,
Nucl.\ Phys.\ B {\bf 517} (1998) 241.


\bibitem{vanNeerven:2000wp}
W.~L.~van Neerven and A.~Vogt,
Phys.\ Lett.\ B {\bf 490} (2000) 111.


\end{thebibliography}
\end{document}